\documentclass[letterpaper, journal]{IEEEtran} %For final style
\usepackage[letterpaper, left=15mm,right=15mm,top=15mm,bottom=15mm]{geometry}

\usepackage{graphicx}
\usepackage{subcaption} %For subfigure
\graphicspath{ {./Figure/} }
\usepackage{multicol} %For single column pict inside two col
\usepackage{array}
\usepackage{amssymb} %For math symbolic
\usepackage{amsmath} %For reference to equations
\usepackage[linesnumbered,ruled,vlined]{algorithm2e}
\usepackage{algpseudocode}% for clean comment
\usepackage{balance}
\usepackage[font=footnotesize]{caption}

\newcommand\blfootnote[1]{%
	\begingroup
	\renewcommand\thefootnote{}\footnote{#1}%
	\addtocounter{footnote}{-1}%
	\endgroup
} %Footnote without number
\usepackage{tikz,xcolor,hyperref}
\definecolor{lime}{HTML}{A6CE39}
\DeclareRobustCommand{\orcidicon}{%
	\begin{tikzpicture}
	\draw[lime, fill=lime] (0,0) 
	circle [radius=0.16] 
	node[white] {{\fontfamily{qag}\selectfont \tiny ID}};
	\draw[white, fill=white] (-0.0625,0.095) 
	circle [radius=0.007];
	\end{tikzpicture}
	\hspace{-2mm}
}

\foreach \x in {A, ..., Z}{%
	\expandafter\xdef\csname orcid\x\endcsname{\noexpand\href{https://orcid.org/\csname orcidauthor\x\endcsname}{\noexpand\orcidicon}}
}
\begin{document}
	\title {\Huge Multiple Antenna Selection and Successive Signal \\Detection for SM-based IRS-Aided Communication}	
	\author{\IEEEauthorblockN{
			Hasan Albinsaid\orcidA{},~\IEEEmembership{Student Member,~IEEE},
			Keshav Singh\orcidB{},~\IEEEmembership{Member,~IEEE},
			Ankur Bansal\orcidC{},~\IEEEmembership{Member,~IEEE},
			Sudip Biswas\orcidD{},~\IEEEmembership{Member,~IEEE},
			Chih-Peng Li\orcidE{},~\IEEEmembership{Fellow,~IEEE},  and
			Zygmunt J. Haas\orcidF{},~\IEEEmembership{Fellow,~IEEE}}% <-this % stops an unwanted space
			\vspace{-1em}
			\thanks{\hrulefill}
			\thanks{This work was supported by the Ministry of Science and Technology of Taiwan under Grants MOST 109- 2218-E-110-006 and MOST 109-2221-E-110-050-MY3.}
			\thanks{H. Albinsaid, K. Singh and C.-P. Li are with the National Sun Yat-sen University, Kaohsiung 80424, Taiwan. (e-mail: hasan@g-mail.nsysu.edu.tw, \{keshav.singh, cpli\}@mail.nsysu.edu.tw).}
			\thanks{A. Bansal is with the Indian Institute of Technology Jammu, Jammu and Kashmir, India. (e-mail: bansal.ankur143@gmail.com).}
			\thanks{S. Biswas is with the Department of ECE, Indian Institute of Information Technology, Guwahati- 781015, India. (email: sudip.biswas@ieee.org).}
			\thanks{Zygmunt J. Haas is with the University of Texas at Dallas, Richardson, TX 75080, U.S.A. (e-mail: haas@utdallas.edu).}
			\thanks{Digital Object Identifier 10.1109/LSP.2021.3071981}	
		}	
	
	\markboth{IEEE SIGNAL PROCESSING LETTERS}%
	{Shell \MakeLowercase{\textit{et al.}}: Bare Demo of IEEEtran.cls for IEEE Communications Society Journals}
	
	\IEEEtitleabstractindextext{%
		\begin{abstract}
			Intelligent reflecting surface (IRS) is being considered as a prospective candidate for next generation wireless communication due to its ability to significantly improve coverage and spectral efficiency by controlling the propagation environment. One of the ways IRS increases spectral efficiency is by adjusting phase shifts to perform passive beamforming. In this letter, we integrate the concept of IRS aided communication to the domain of multi-direction beamforming, whereby multiple receive antennas are selected to convey more information bits than existing spatial modulation (SM) techniques at any specific time. To complement this system, we also propose a successive signal detection (SSD) technique at the receiver. Numerical results show that the proposed design is able to improve the average successful bits transmitted  (ASBT) by the system, which outperforms other state-of-the-art methods proposed in literature.
		\end{abstract}
	\vspace{-1em}
	}
	
	% make the title area
	\maketitle
	\IEEEoverridecommandlockouts
	\IEEEpubid{
		\begin{minipage}{\textwidth}\ \\ \\ \\ \\[12pt]
			\centering
			1558-2558~\copyright~2021 IEEE. Personal use is permitted, but republication redistribution requires IEEE permission. \\ See https://www.ieee.org/publications/rights/index.html for more information.
		\end{minipage}
	}
	\IEEEdisplaynontitleabstractindextext
	\IEEEpeerreviewmaketitle

	\section{Introduction}
	\IEEEPARstart{W}{ith} 5G being deployed in phases in various parts of the world, the focus has shifted towards developing the post-5G wireless technologies with futuristic networking trends \cite{di2019smart}. New applications and user requirements demand new communication paradigms in the physical layer~\cite{WSaad2020, raghavan2019evolution}. Towards this goal, intelligent reflecting surface (IRS) assisted communication has gained immense popularity recently~\cite{nadeem2019intelligent} that can significantly improve coverage and spectral efficiency by controlling the wireless propagation environment. 
	In particular, IRS consists of a large number of reconfigurable passive and low-cost reflecting elements that are able to reflect the incoming signal in the desired direction by adjusting the phase shifts of its elements~\cite{liaskos2018new,taha2019enabling,you2020wireless}.% E.g., an array of UAVs.
	
	In recent developments on IRS, passive beamforming, and information transfer via IRS have been proposed in~\cite{basar2020reconfigurable,wu2018intelligent}, which show that IRS devices can aid transmission of information bits by adopting single antenna selection (SAS) spatial shift keying (SSK) and spatial modulation (SM) at the receiver. However, these techniques are limited by the number of bits that can be conveyed through both antenna indices or the used $M$-ary digital modulation scheme. 
	Accordingly, to overcome the limitations of literature in a related topic, we propose a technique by adopting multiple antenna selections (MAS) at the receiver and using $M$-ary digital modulation at the transmitter that enables the system to carry more information bits in comparison to existing SAS-SSK or SAS-SM. However, due to the increasing bits conveyed by this method, it becomes challenging to decode the information at the receiver, to which end we propose a novel signal detection technique. The studied system in our paper may share similarities to other IRS and SM-based works, both in terms of the system concept, service requirements, and design guidelines. However, the selection of multiple antennas in SM and multiple modulations introduces fundamental new challenges in signal detection for the IRS system, which is the main focus of this paper. The primary contributions are summarized as:	
	\begin{itemize}
		\item We propose a design that amalgamates IRS and SM and eliminates the drawback in the throughput calculation with base-$2$ logarithm of the number of receive antennas by selecting multiple antennas at the receiver and $M$-ary digital modulation at the transmitter.
		\item We adopt the concept of superposition coding (SC) to combine $M$-ary digital modulation symbols into a single complex-valued scalar. As the transmitter only transmits a single complex-valued, we propose successive signal detection (SSD), wherein the detection process is performed sequentially to extract the information received from the transmitter.
		\item Using numerical results, we show that the proposed design improves the average successful bits transmitted (ASBT) {which is defined as the average number of bits that can be successfully transmitted in each transmission}. Moreover, IRS's impact reflecting elements on bit error rate (BER) performance is also studied.
	\end{itemize}

\section{System Model}
\begin{figure*}[t!]
	\centering
	\includegraphics[width=0.65\textwidth]{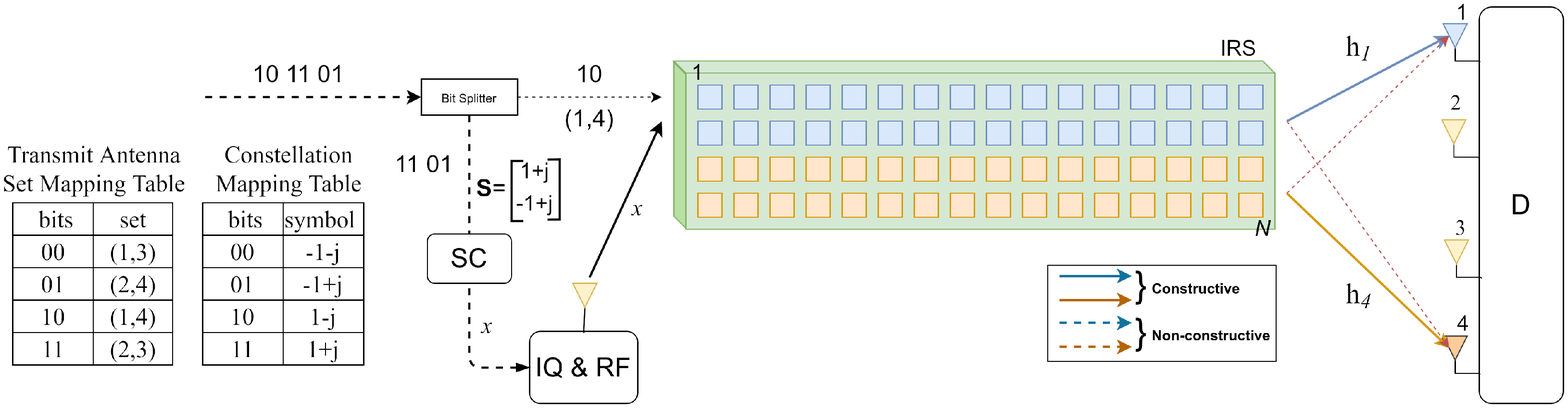}
	\caption{IRS-assisted communication system with $N_{r}=4$ and $N_{p}=2$.}
	\label{fig:LIS_SIC}
\end{figure*}
\begin{figure}[t!]
	\centering
	\includegraphics[scale=0.4]{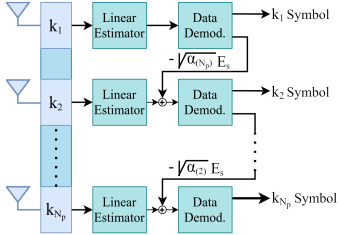}
	\caption{SSD process on selected receive antennas.}
	\label{fig:SIC_ilustration}
\end{figure}
The proposed IRS-assisted communication system is shown in Fig.\ref{fig:LIS_SIC}, which consists of an RF source and a destination ($D$) equipped with $N_r$ receive antennas. The communication is established through an IRS aid, comprising of $N$ low-cost passive reflector elements that reflect the incoming information bits to $D$. Among $N_r$ possible receive antennas, $N_p$ receive antennas are selected as target antennas with $N_p \ll N_r$. Therefore the total possibility of selected receive antennas is given by $N_r \choose N_p$, where $N_r \choose N_p$ represents the binomial coefficient. Further, only $C=2^{\lfloor \mathrm{log}_{2} \binom{N_{r}}{N_{p}}\rfloor}$ receive antenna combinations (RAC) are permitted to be legitimate combinations and the remaining combinations are considered as illegitimate, where $\lfloor \cdot \rfloor$ denotes the floor operation. The RAC matrix is denoted by $\mathbf{R}\in \mathbb{N}^{C\times N_p}$.

{In each time slot, the vector of information bits $\mathbf{b}$ will be divided into two-part, i.e., RAC modulated bits and symbols modulated bits using quadrature amplitude modulation (QAM) as $L_1$ and $L_2$, respectively. The RAC modulation part is conveyed to the selected receive antennas' indices with $L_1=\lfloor \mathrm{log}_{2} \binom{N_{r}}{N_{p}}\rfloor$ bits, and the symbol modulation part conveys $N_p$ of \textit{M}-ary symbol modulation $\mathbf{s}_i$ from the set of constellation $\mathcal{S}=[\mathcal{S}_1,\dots,\mathcal{S}_M]$, with $L_2=N_p\mathrm{log}_2M=N_p\mu$ bits, where $i\in\{1,\dots,N_p\}$.}
		
Let $\mathbf{H} \in \mathbb{C}^{N_r \times N}$ denote the wireless fading channel matrix between IRS and $D$, whose each entry follows a complex Gaussian distribution with zero-mean and unit-variance. {We assume perfect channel state information (CSI) at both nodes. Since the channel matrix $\mathbf{H}$ is a complex-valued matrix, it has an amplitude and a phase for each element that can be expressed as $h_{m,n}=\beta_{m,n}e^{j\psi_{m,n}}$ for $m\in\{1,\dots,N_r\}$ and  $n\in\{1,\dots,N\}$}. 
\blfootnote{\textit{Notation}: In the following, uppercase and lowercase bold letters represent matrices and vectors, respectively. $\mathbf{A}[m,n]$ represent \textit{m}-th row and \textit{n}-th column of matrix $\mathbf{A}$. $\mathbf{a}_m$ denotes the \textit{m}-th element of vector $\mathbf{a}$.} We assume that the transmitter is equipped with a single transmit antenna. In order to transmit $N_p$ independent symbols to $D$ which has $N_p$ selected receive antennas, in each time slot we implement SC~\cite{yang2020intelligent} at the transmitter side and SSD at the receiver side. {The idea behind SC is to combine all \textit{M}-ary symbols into a single complex-valued scalar by multiplying the ratio value $\boldsymbol{\alpha}_i$ for each symbol, where $\Sigma_{i=1}^{N_p}\boldsymbol{\alpha}_i=1$ and $\boldsymbol{\alpha}_i<\boldsymbol{\alpha}_{i+1}$. }
	
Next, let $p$ denote the index set of RAC, where $p\in\{1,2,\dots,C\}$. From the $p$-th index set of RAC, we have to sort  the weight of the row target channel-matrix $\tilde{\mathbf{H}}\in \mathbb{C}^{N_p \times N}$, where $\tilde{\mathbf{H}}=\mathbf{H}[\mathbf{l}_{p},:]$ and $\mathbf{l}_{p}=\mathbf{R}[p,:]$. The weight of row target channel-matrix $\tilde{\mathbf{H}}$ can be written as $\mathbf{w}_i=\|\tilde{\mathbf{h}}_{p_i}\|_2$, where $\tilde{\mathbf{h}}_{p_i}$ denotes the $i$-th row of $\tilde{\mathbf{H}}$. To achieve fairness in power allocation, the weight is sorted in descending order and is expressed as
\begin{equation} \label{eq_w}
[k_1, k_2,\dots,k_{N_p}]=\mathrm{arg \,sortd}(\mathbf{w})\,.
\end{equation}
{By using the sorted channel condition, we can perform SC to all transmit symbol with a fair ratio that can be expressed as 
\begin{equation} \label{eq_x}
x=\sum\nolimits_{i=1}^{N_p}\sqrt{\boldsymbol{\alpha}_i}E_s\mathbf{s}_{k_i}\,,
\end{equation}
where $E_s$ is symbol's energy and $\mathbf{s} \in \mathbb{C}^{N_p \times 1}$.}

{As stated earlier, since the IRS consists of $N$ reflecting elements, it can maximize the received SNR at the specific RAC by adjusting the reflecting phases $\{\phi_n\}_{n=1}^N$ according to the information bits. For this purpose, the first $L_1$ information bits specify the index $p$ of RAC, and then the IRS adjusts its phase according to it. Since there are $N_p$ selected receive antennas in each time slot, $N$ reflector elements of IRS are subdivided into $N_{p}$ parts such that the receive antennas have equal amplification. Thus, we have
	\begin{equation} \label{eq_phi}
	\phi_{(i-1)\Delta+1:i\Delta}=\psi[\mathbf{R}[p,i],(i-1)\Delta+1:i\Delta]\,,
	\end{equation}
	where $\Delta=\lfloor \frac{N}{N_p} \rfloor$ is the number of reflectors dedicated for each selected receive antenna and $\psi[m,n]$ denotes the phase of $\mathbf{H}$ in $m$-th row and $n$-th column. By adopting this approach, we obtain the reflector phase vector at the \textit{p}-th index set of RAC  $\boldsymbol{\theta}^{(\mathbf{l}_{p})}\in \mathbb{C}^{N\times 1}$, which is given as 
	\begin{equation} \label{eq_theta}
	\boldsymbol{\theta}^{(\mathbf{l}_{p})}_{(i-1)\Delta+1:i\Delta}\!=\!\boldsymbol{\exp}(\!-j\;\mathbf{phase}(\tilde{\mathbf{H}}[i,(i\!-\!1)\Delta\!+\!1:i\Delta]))\,.
	\end{equation}
	}
	The method for transmission of superposition signal and to obtain the reflector phase vector is summarized in Algorithm~\ref{alg:transmit}.
	 Now, the received signal vector $\mathbf{y} \in \mathbb{C}^{N_r \times 1}$ can be given as  
	\begin{equation} \label{eq_y}
	\mathbf{y}=\mathbf{H}\boldsymbol{\theta}^{(\mathbf{l}_{p})}x+\mathbf{n}\,,
	\end{equation}
	where $\mathbf{n}\in\mathbb{C}^{N_r\times1}$ is the additive noise vector whose entries follow a complex Gaussian distribution with zero mean and variance $\sigma^2$. {Multiplication of $\mathbf{H}$ with $\boldsymbol{\theta}^{(\mathbf{l}_{p})}$ resulting in constructive and non-constructive parts, hence the received signal for $i$-th selected receive antenna is given by 
	\begin{equation}\label{eq_y1}
	\begin{split}
	&\mathbf{y}_i=\underbrace{\sum\nolimits_{n=(i-1)\Delta+1}^{i\Delta}\boldsymbol{\beta}_{p_i,n}x}_\text{Constructive part} \,+ \\
	&\underbrace{\sum\nolimits_{q=1, q\neq i}^{N_p}\sum\nolimits_{\;n=(q-1)\Delta+1}^{q\Delta}\boldsymbol{\beta}_{p_i,n}e^{j\;(\boldsymbol{\phi}_{n}-\boldsymbol{\psi}_{p_i,n})}x}_\text{Non-constructive part}+\mathbf{n}_i\,.
	\end{split}
	\end{equation}}

	Since reflectors amplify the signal at the selected receive antennas, it results in the maximizing instantaneous SNR as 
	\begin{equation} \label{eq_snrmax}
	\begin{split}
	\mathrm{max}\{\gamma\}=E_s\sum\nolimits_{i=1}^{N_p}\Big(\frac{\mid \sum_{n=(i-1)\Delta+1}^{i\Delta}\boldsymbol{\beta}_{p_i,n}\mid ^2}{\sigma^2} \\+ \frac{\mid \sum_{q=1, q\neq i}^{N_p} \sum_{n=(q-1)\Delta+1}^{q\Delta}\boldsymbol{\beta}_{p_i,n}e^{j(\boldsymbol{\phi}_n-\boldsymbol{\psi}_{p_i,n})} \mid ^2}{\sigma^2}\Big)\,,
	\end{split}
	\end{equation}where $\gamma$ is defined as 
	\begin{equation} \label{eq_snr}
	\gamma=E_sN_p\frac{\mid \sum_{q=1}^{N_p} \sum_{n=(q-1)\Delta+1}^{q\Delta}\boldsymbol{\beta}_{p_q,n}e^{-j\boldsymbol{\psi}_{p_q,n}} \mid ^2}{\sigma^2}.
	\end{equation}
	From (\ref{eq_y}), the optimal ML-based demodulator is given by 
	\begin{equation} \label{eq_ML}
	(\hat{p},\hat{x})= \mbox{arg} \min_{p,x\in\mathbb{X}} {\| \mathbf{y}-\mathbf{H}\boldsymbol{\theta}^{(\mathbf{l}_{p})}x\|}_F^{2}\,,
	\end{equation}
	where $\mathbb{X}=[x_1,x_2,\dots,x_\tau]$ is the set of all possibilities of $N_p$ superposition of \textit{M}-QAM symbol modulation,  $\tau=\mu^{N_p}$, and $\|\cdot\|_F$ denotes the Frobenius norm of a matrix.
	It is apparent that the ML-based demodulator algorithm jointly detects the selected receive antennas and digital modulation by exhaustive search from all possibilities on the transmitted signal vector $\mathbf{b}$. However, it may have a prohibitive computational complexity, especially for the larger number of $N_r$ and $N_p$.
	\SetKwInput{KwInput}{Input} % Set the Input
	\SetKwInput{KwOutput}{Output} % set the Output
	\begin{algorithm}[t!]
		%\DontPrintSemicolon
		\KwInput{$\mathbf{H}$, $\boldsymbol{\alpha}$, $\mathbf{R}$, $N_p$, $\mathbf{b}$, $L_1$, $L_2$, $\mu$, $\Delta$}
		$p=\mathbf{bin2dec}(\mathbf{b}_{1:L_1})$
		
		$\mathbf{l}_{p}=\mathbf{R}[p,:]$
		
		$\tilde{\mathbf{H}}=\mathbf{H}[\mathbf{l}_{p},:]$
		
		$\mathbf{w}_i=\| \tilde{\mathbf{h}}_{p_i} \|_2 \qquad$ with $i\in \{1,\dots,N_p\}$ 
		
		$[k_1,k_2,\dots,k_{N_p}]=\mathrm{arg\,sortd} (\mathbf{w})$
		
		$x=0$
		
		\For{$i\gets1$ to $N_p$ by $1$}{
			$x=x+\sqrt{\boldsymbol{\alpha}_i}E_s\mathcal{S}_{\mathbf{bin2dec}(\mathbf{b}[L_1+({k_i}-1)*\mu+1:L_1+{k_i}*\mu])}$
		}
		
		\For{$i\gets1$ to $N_p$ by $1$}{
			$\mathbf{r}=[(i-1)\Delta+1:i\Delta]$
			
			$\boldsymbol{\theta}^{(\mathbf{l}_{p})}_{\mathbf{r}}=\boldsymbol{\exp}(-j\; \mathbf{phase}(\tilde{\mathbf{H}}[i,\mathbf{r}]))$
		}

		\KwOutput{Superposition signal=$x$; Reflector phase=$\boldsymbol{\theta}^{(\mathbf{l}_{p})}$ 
		}
		\caption{Algorithm for transmission of superposition signal and to obtain the reflector phase vector}
		\label{alg:transmit}
	\end{algorithm}

	\SetKwInput{KwInput}{Input}                % Set the Input
	\SetKwInput{KwOutput}{Output}              % set the Output
	\begin{algorithm}[t!]
		%\DontPrintSemicolon
		\KwInput{$\mathbf{y}$, $\mathbf{H}$, $\mathbf{I}$, $\mathbf{R}$, $N$, $N_p$, $E_s$, $n_c$, $\Lambda$, $\Delta$, D=$\infty$}
		
		\For{$m\gets1$ to $N_r$ by $1$}{
			$\mathbf{a}_{m}=\mid \mathbf{y}_m \mid ^2$
		}
		$(\hat{\mathbf{a}})=\mathrm{arg\,sortd}(\mathbf{a})$
		
		$\mathbf{\hat{R}}=$ all $\mathbf{R}$ that's containing the first $n_{c}$ scalar value of vector $\hat{\mathbf{a}}$
		
		\For{$u\gets1$ to $\lambda$ by $1$}{
			$\mathbf{z}_{u}=\sum_{i=1}^{N_p} \mid \mathbf{y}_{\mathchoice{}{}{\scriptscriptstyle}{}\mathbf{\hat{R}}[u,i]} \mid ^2$
		}
		$(\hat{\mathbf{c}})=\mathrm{arg\,sortd}(\mathbf{z})$

		\For{$v\gets1$ to $\Lambda$ by $1$}{
			$\hat{p}=$row index of RAC where $\mathbf{R} = \mathbf{\hat{R}}[\hat{\mathbf{c}}_v,:]$
			
			$\hat{\mathbf{H}}=\mathbf{H}[\mathbf{l}_{\hat{p}},:]$
			
			$\hat{\mathbf{w}}_i=\| \hat{\mathbf{h}}_{p_i} \|_2 \qquad$ with $i\in \{1,\dots,N_p\}$
			
			$[k_1,k_2,\dots,k_{N_p}]=\mathrm{arg\,sorta}(\hat{\mathbf{w}})$
			
			\For{$i\gets1$ to $N_p$ by $1$}{
				$\mathbf{r}=[(i-1)\Delta+1:i\Delta]$
				
				$\boldsymbol{\theta}^{(\mathbf{l}_{\hat{p}})}_{\mathbf{r}}=\boldsymbol{\exp}(-j\; \mathbf{phase}(\hat{\mathbf{H}}[i,\mathbf{r}]))$
			}
			
			$\ddot{\mathbf{s}}_{k_{1}}=\mathbb{Q}\big(\frac{\mathbf{y}_{\mathbf{R}}\mathchoice{}{}{\scriptscriptstyle}{}[\hat{p},k_{1}]}{\hat{\mathbf{H}}[k_{1},:]\boldsymbol{\theta}}\big)$
			
			\For{$i\gets2$ to $N_p$ by $1$}{
				$\ddot{\mathbf{s}}_{k_{i}}=\mathbb{Q}\big(\frac{\mathbf{y}_{\mathbf{R}}\mathchoice{}{}{\scriptscriptstyle}{}[\hat{p},{k_{i}}]}{\hat{\mathbf{H}}[{k_{i}},:]\boldsymbol{\theta}}-\sqrt{\boldsymbol{\alpha}_{(N_p-i+2)}}E_s\ddot{\mathbf{s}}_{k_{i-1}} \big)$
			}
			
			$\hat{x}=0$
			
			\For{$i\gets1$ to $N_p$ by $1$}{
				$\hat{x}=\hat{x}+\sqrt{\boldsymbol{\alpha}_{(N_p-i+1)}}E_s\ddot{\mathbf{s}}_{k_{i}}$
			}
			
			$d= {\| \mathbf{y}-\mathbf{H}\boldsymbol{\theta}^{(\mathbf{l}_{\hat{p}})}\hat{x}\|}_F^{2}$
			
			\If{$d<D$}{
				$D=d$
				
				$\hat{\mathbf{s}}=\ddot{\mathbf{s}}$
				
				$\mathbf{l}_{\hat{p}}=\mathbf{R}[\hat{p},:]$
			}
		}

		\KwOutput{Target Antennas=$\mathbf{l}_{\hat{p}}$, Digital modulation=$\hat{\mathbf{s}}$	}
		\caption{Proposed IRS MAS-SSD Algorithm}
		\vspace{-0.5em}
		\label{alg:decoder}
	\end{algorithm}

	\section{Proposed IRS-SSD Detector}
	Since the optimal ML-based detector seeks all possibilities on the transmitted signal vector $\textbf{b}$, we have to reduce the computational complexity of this detector to make it feasible for practical applications. One way to reduce the computational complexity is by sorting the most likely combinations of selected receive antennas, followed by calculating the euclidean distance and choosing the combination with the smallest euclidean distance. The details of this algorithm are depicted below and summarized in Algorithm~\ref{alg:decoder}.
	\subsection{RAC sorter}
	First, we use the greedy detector approach to look for the antennas with the highest amplitude, it can be expressed as
	\begin{equation} \label{eq_ahat}
	\hat{\mathbf{a}}=\operatorname*{arg\,sortd}_{m\in \{1,\dots,N_r\}}(\mid \mathbf{y}_m \mid ^2)\,,
	\end{equation}
	where $\hat{\mathbf{a}}\in \mathbb{N}^{N_r \times 1}$ is the argument vector of sorted amplitude $\mathbf{y}$ (from highest to lowest).
	Now, to reduce the number of RAC candidates we have, let us take the first $n_c$ highest value of vector $\hat{\mathbf{a}}$ and create a new RAC containing it. As a result, we have $\mathbf{\hat{R}}\in\mathbb{N}^{\lambda \times N_p}$ as our new RAC matrix, where $\lambda$ is the number of candidates left after the previous process.	
	
	Since we aim to select $N_p$ antennas simultaneously in each time slot, we compute $\lambda$ weight combinations of selected receive antennas and determine which combination has the highest probability to be a solution, yielding
	\begin{equation} \label{eq_zu}
	\mathbf{z}_{u}=\sum\nolimits_{i=1}^{N_p} \mid \mathbf{y}_{\mathchoice{}{}{\scriptscriptstyle}{}\mathbf{\hat{R}}[u,i]} \mid ^2,
	\end{equation}
	where $u\in\{1,\dots,\lambda\}$. Next, we sort the amplitude of the highest probability RAC index as  
	\begin{equation} \label{eq_chat}
	\hat{\mathbf{c}}=\mathrm{arg\,sortd}\;(\mathbf{z})\,.
	\end{equation}
	Now, we get sorted $\lambda$ set of new RAC given by $\mathbf{\hat{R}}[\hat{\mathbf{c}}_u,:]$, where smaller $u$ denotes the most likely solution of RAC index.

	\begin{figure*}[t!]
		\centering
		\includegraphics[trim=170 0 120 0,width=0.75\textwidth, height=11em]{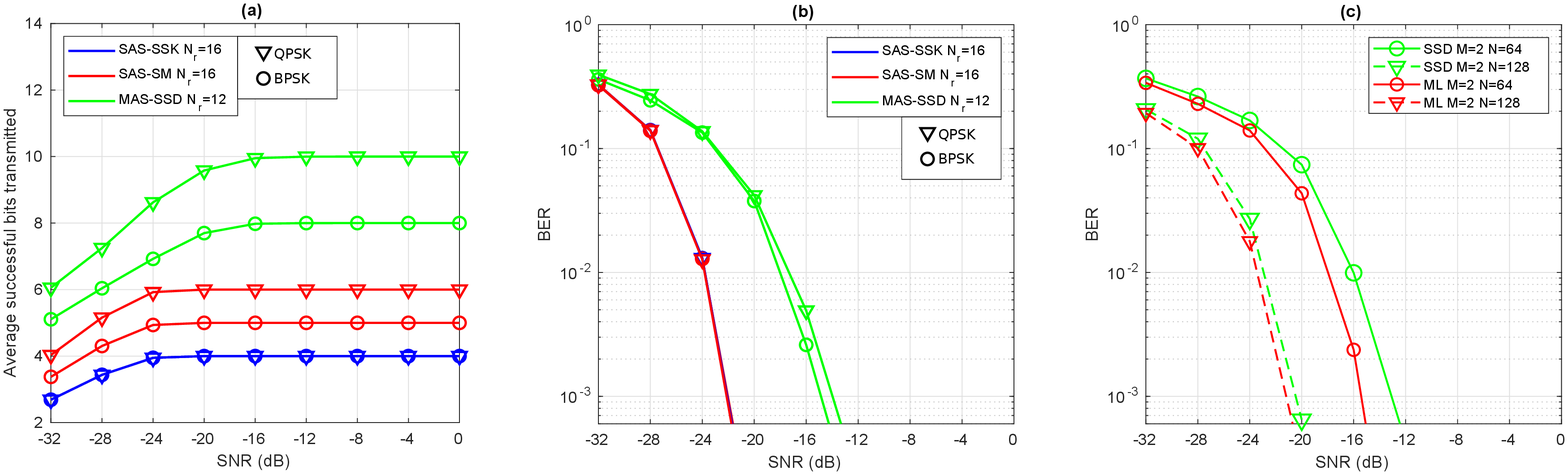}
		\caption{{(a) ASBT and  (b) BER comparison for the proposed scheme with SAS-SSK and SAS-SM schemes under BPSK modulation and QPSK modulation.} (c) BER comparison for the proposed scheme with MAS-ML under various number of $N$}
		\label{fig:Results}
	\end{figure*}

	\vspace{-0.5em}
	\subsection{SSD decoder}
	Since our primary goal is to reduce the computational complexity, let we consider $\Lambda$ as given number of iteration for the decoder, where  $\Lambda\leqslant\lambda$. Let $\hat{p}$ be a row index of RAC given as $\mathbf{R}[\hat{p},:] = \mathbf{\hat{R}}[\hat{\mathbf{c}}_v,:]$, where $v\in\{1,\dots,\Lambda\}$. For $\hat{p}$-th index of RAC, we select a row channel $\hat{\mathbf{H}}$, where $\hat{\mathbf{H}}=\mathbf{H}[\mathbf{l}_{\hat{p}},:]$. 
	
	Similar to the SC process, we have to calculate the reflection phase as (\ref{eq_theta}) and sort the weights of selected row channel $\hat{\mathbf{H}}$ in ascending order as 
	\begin{equation} \label{eq_w2}
	[k_1, k_2,\dots,k_{N_p}]=\mathrm{arg\,sorta}(\hat{\mathbf{w}})\,,
	\end{equation}where $\hat{\mathbf{w}}_i=\|\mathbf{\hat{h}}_{\hat{p}_i}\|_2$.
	
	The SSD process uses the result of~\eqref{eq_w2} to predict the transmitted symbol from the transmitter, as illustrated in Fig. \ref{fig:SIC_ilustration}, where the symbols are decoded from the weakest to the strongest channel. Since non-constructive signals appear in each antenna, we adopt SSD to mitigate the non-constructive part of the signal except for the $k_1$ because it has the highest value multiplicative ratio $\boldsymbol{\alpha}$. As a result, we get $\ddot{\mathbf{s}}\in \mathbb{c}^{N_p \times 1}$ as the predicted symbol. The detailed process about the SSD is shown in Algorithm \ref{alg:decoder} (line 16-18). {To check whether the candidate we choose is correct or not, we have to calculate the euclidean distance between the received signal and the prediction $\Lambda$ times and looking for the minimum distance. Since the transmitted signal is in the form of SC, the symbol $\ddot{\mathbf{s}}$ has to be converted to SC form, which can be expressed as 
	\begin{equation} \label{eq_xhat}
	\hat{x}=\sum\nolimits_{i=1}^{N_p}\sqrt{\boldsymbol{\alpha}_{(N_p-i+1)}}E_s\ddot{\mathbf{s}}_{k_{i}}.
	\end{equation}}
	Then, the euclidean distance can be obtained by 
	\begin{equation} \label{eq_distance}
	d= {\| \mathbf{y}-\mathbf{H}\boldsymbol{\theta}^{(\mathbf{l}_{\hat{p}})}\hat{x}\|}_F^{2}.
	\end{equation}
	
	Since we have $\Lambda$ candidates as the solutions, it is required to compare the results in terms of euclidean distance $d$ for each $\Lambda$ and select the smallest one shown in  Algorithm \ref{alg:decoder} (line 23-26). Finally, we obtain the predicted RAC $\mathbf{l}_{\hat{p}}$ and the predicted signal-vector $\hat{\mathbf{s}}$ as the solutions.
	
	The complexity for the SSD detector in terms of multiply-and-accumulate (MAC) operations is given as 
	\begin{equation} \label{C_SSD}
	C_{\mathchoice{}{}{\scriptscriptstyle}{}\mathbf{SSD}}=\Lambda(8N_{r}N+10N_{r}-1)+\lambda(N_{p}-1)+3N_{r}\,,
	\end{equation}
		while the complexity for the ML detector according to (\ref{eq_ML}) is computed as 
	\begin{equation} \label{C_ML}
	C_{\mathchoice{}{}{\scriptscriptstyle}{}\mathbf{ML}}=2^{\mathchoice{}{}{\scriptscriptstyle}{}\left(L_{1}+L_{2}\right)}(8N_{r}N+10N_{r}-1)\,.
	\end{equation}

\section{Simulation Results} 
This section numerically investigates the two algorithms' performance for the proposed IRS with implementing MAS and SSD detector at the receiver. We use 12 antennas at the receiver with two selected antennas at any particular state. We compare our design's performance to \cite{basar2020reconfigurable}, which uses 16 antennas at the receiver where one antenna is selected. Some other parameters used are as follows:  $N=64$, $\boldsymbol{\alpha}=[0.2,0.8]$, $n_c=6$, $\Lambda=8$, and number of data stream $N_s=10^5$.  
Let us begin with the ASBT performance that shows the average bit length value for each signal transmission in Fig. \ref{fig:Results}(a). It can be seen that the proposed design with MAS-SSD outperforms both SAS-SSK and SAS-SM even with a lesser number of receive antennas. It can be explained by the fact that for BPSK modulation, SAS-SSK can convey 4 bit SSK, SAS-SM can convey 4 bit SSK and 1 bit digital modulation. On the other hand, MAS-SSD can convey 6 bit SSK and 2 bit digital modulation. For QPSK modulation, SAS-SSK can convey 4 bit SSK, SAS-SM can convey 4 bit SSK and 2 bit digital modulation. On the contrary, MAS-SSD can convey 6 bit SSK and 4 bit digital modulation. It is worth noting that the proposed method has the ability to carry more information with a lesser number of receive antennas than other contemporary techniques.  

In Fig.~\ref{fig:Results}(b), we use 12 antennas for the proposed MAS-SSD scheme instead of 16 antennas because $\lfloor \mathrm{log}_{2} \binom{12}{2}\rfloor$ is equal to $\lfloor \mathrm{log}_{2} \binom{16}{2}\rfloor$ and thus, this reduces the number of receive antennas but with the same result. From Fig. 3(a) and Fig.~\ref{fig:Results}(b), we find that our method can be a solution to increase the number of transmitted bits at the price of increasing BER.

Finally, in Fig. \ref{fig:Results}(c) we show the quality of the received signal with MAS under BPSK modulation in terms of BER with varying number of reflector elements, i.e., $N=64$ and $N=128$. It can be seen from the figure that as the number of passive reflector-elements increases, the BER performance of the IRS-based network becomes better, and the SSD detector still reasonable compared to the ML detector.

\begin{table}[t!]
	\renewcommand{\arraystretch}{0.65}
	\begin{center}
		\caption{Notation List}
		\label{tab:table1}
		\begin{tabular}{m{1.6cm}|>{\arraybackslash}m{5.25cm}}
			\hline
			\textbf{Symbol} & \textbf{Meaning}\\ % <-- added & and content for each column
			\hline
			\hline
			$\mathbf{exp}(\bullet)$ & exponential function  \\ % <--
			\hline
			$\mathbf{phase}(\bullet)$ & function for convert complex-valued into an phase in radian  \\ % <--
			\hline
			$\mathbf{bin2dec}(\bullet)$ & function to convert from binary to decimal  \\ % <--
			\hline
			$\mathbb{Q}(\bullet)$ & Digital demodulation function \\ % <--
			\hline
		\end{tabular}
	\end{center} 
\end{table}	

\section{Conclusion} 
A multi-direction beamforming with MAS at the receiver was studied in this work for an IRS-aided communication system. In line with the compact transmitter pre-processing, an SSD detector was also proposed to translate the received information. Numerical results demonstrated the effectiveness of the proposed method, which is able to convey more information bits than existing SAS-SM techniques with effective BER performance. The proposed IRS design with MAS-SSD meets the criteria of the call for unconventional wireless communications that can help to meet new networking demands beyond 5G, such as 6G wireless networks.
	\balance
	\bibliographystyle{IEEEtran}
	\bibliography{IEEEabrv,bibliography_ex}	
\end{document}